# Complete 'melting' of charge order in hydrothermally grown $Pr_{0.57}Ca_{0.41}Ba_{0.02}MnO_3$ nanowires

K.N.Anuradha, S.S.Rao and S.V.Bhat[*]
*Department of Physics, Indian Institute of Science, Bangalore, India*

Nanowires of $Pr_{0.57}Ca_{0.41}Ba_{0.02}MnO_3$ (PCBM) (diameter ~ 80-90 nm and length ~ 3.5 μm) were synthesized by a low reaction temperature hydrothermal method. Single-phase nature of the sample was confirmed by XRD experiments. Scanning electron microscopy (SEM) and transmission electron microscopy (TEM) were used to characterize the morphology and microstructures of the nanowires. While the bulk PCBM is known to exhibit charge order (CO) below 230 K along with a ferromagnetic transition at 110 K, SQUID measurements on the nanowires of PCBM show that the charge order is completely absent and a ferromagnetic transition occurs at 115 K. However, the magnetization in the nanowires is observed to be less compared to that in the bulk. This observation of the complete 'melting' of the charge order in the PCBM nanowires is particularly significant in view of the observation of only a weakening of the CO in the nanowires of $Pr_{0.5}Ca_{0.5}MnO_3$. Electron paramagnetic resonance experiments were also carried out on the PCBM nanowires using an X-band EPR spectrometer. Characteristic differences were observed in the line width of nanowires when compared with that of the bulk.

**1. Introduction:**

The rare earth manganites $R_{1-x} A_x MnO_3$ (where R is a trivalent rare earth ion and A is a divalent alkaline earth ion) are continuing to receive intense experimental and theoretical attention [1,2]. This interest is mainly driven by their complex phase diagrams with fragile phase boundaries resulting from a delicate balance between various interactions such as (double) exchange, super exchange, Jahn-Teller and crystal field effects [3,4]. These materials are characterized by a competition between and the coexistence of the colossal magneto resistive (CMR) ferromagnetic (FM) and charge ordering (CO), antiferromagnetic ground states. It is generally believed that the two phases are mutually exclusive; e.g. in the prototype manganite $La_{1-x}Ca_xMnO_3$ for $0.2 < x < 0.5$ the material is ferromagnetic and metallic while for $0.5 < x < 0.9$ it is antiferromagnetic and charge ordered [5]. Examples of the same material showing both CO and FM phases such as $Nd_{0.5}Sr_{0.5}MnO_3$ [6] are rare. It was also observed that minute perturbations of the system with magnetic fields of the order of a few Tesla, electric field and radiation could cause a "melting" of the CO phase and the appearance of an FM phase [7]. While the initial expectations about the possible technological applications of the CMR manganites are not fulfilled, it is felt that nanometric magnetic structures of manganites might still lead to useful devices.

Recently a few studies on nanometric manganites have been published which give an indication of interesting effects of the reduction in size in these systems though many results are contradictory of each other. For example, while some reports [8] have concluded that the ferromagnetic transition temperature $T_C$ decreases with a reduction in the particle size, others see no change [9] or even an increase [10]! An increase in the resistivity with the decreasing size has been observed [11] by some workers whereas others see a decrease [12]. It is possible that these different effects originate in the predominance of different intrinsic effects such as grain boundary contribution, strain, lattice distortion, changes in the bond lengths and bond angles in different systems but one cannot completely rule out the influence of the preparation techniques/conditions and other extrinsic causes. Therefore it is essential to investigate many more systems to find out if certain commonalities are present behind these diverse observations.

In contrast to the studies on the CMR compounds referred to above, there is hardly any report on charge ordering nano manganites except on nanowires of $Pr_{0.5}Ca_{0.5}MnO_3$ [13] and nanoparticles of $Nd_{0.5}Ca_{0.5}MnO_3$ [14] and $Pr_{0.5}Sr_{0.5}MnO_3$ [15]. In the nanowire sample only a partial suppression of CO is observed whereas in the nanoparticles of $Nd_{0.5}Ca_{0.5}MnO_3$ the CO is completely absent. This raises the question about the incomplete suppression of the CO in the nanowires: is this a consequence of the material being microscopic in one dimension? Is it necessary to have a 3-dimensional nano material to have full suppression the charge order? In the present work we attempt to provide an answer to this question.

$Pr_{0.57}Ca_{0.43}MnO_3$ is a charge ordering system with $T_{CO}$ = 230 K. Recently Zhu et al., have shown [16] that low (~2%) doping of Ba for Ca has a profound effect on the phase diagram, making the material go nearly fully ferromagnetic around 2.5 K. We have confirmed this result by magnetization and electron paramagnetic resonance studies[17] and find that this system provides one of the rare examples of a charge ordering but ferromagnetic (with the two transition temperatures being widely different) material. In this work, we investigate the effect of reducing the size in two dimensions i.e. we prepare nanowires of $Pr_{0.57}Ca_{0.41}Ba_{0.02}MnO_3$ and study its properties. Quite significantly we find that the CO phase is completely absent in the nanowires, in contrast with the nanowires of $Pr_{0.5}Ca_{0.5}MnO_3$ showing that it is not necessary to have all the three dimensions nanometric for the charge order to completely disappear. To the best of our knowledge, this is the first report of complete melting of charge order in a nanowire.

**2. Experimental:**

Stoichiometric proportions of high purity chemicals, $Pr(NO_3)_3 \cdot 6H_2O$, $Ca(NO_3)_2 \cdot 4H_2O$, $Ba(NO_3)_2$, $KMnO_4$, $MnCl_2 \cdot 4H_2O$ and KOH were dissolved in de-ionized water. Alkalinity of the solution was adjusted to a pH of 14. The solution was then transferred into a teflon vessel. The vessel was placed in a stainless steel container and sealed, and was heated in a furnace at 270°C for 50 hours. After the heat treatment the autoclave was cooled and depressurized. The final product was washed with de-ionized water and dried in air at 100°C. The composition of the sample was checked by EDAX. The sample was further characterized by XRD, SEM, and TEM. The temperature dependence of the magnetization of the nanowires was recorded using SQUID magnetometry in the temperature range of 10-300 K. Electron magnetic resonance (EMR) signals were recorded from the nanowires and the bulk polycrystalline sample at room temperature in the X-band region.

## 3. Results and discussion:

Figure 1 shows the powder XRD pattern of the sample recorded at room temperature in the 2θ scan range 10° –100° at a scan rate of 0.01°/10sec. All the peaks could be indexed to the orthorhombic space group Pbnm and the Rietveld refinement (Fig.1, solid line: $R_w$ = 8.5 %) resulted in the lattice parameters  a = 5.4572 A°, b = 7.6326 A°, c = 5.4474 A° and the unit cell volume V= 226.8985 A°$^3$. As expected, the X-ray diffraction peaks in the nanowires were observed to be broader than those of the bulk manganite. These parameters indicate increases in the a and the c parameters as well as the volume of the unit cell but a small contraction of the b parameter compared to the values in the bulk (a = 5.4229 A°, b = 7.6336 A°, and c = 5.4078 A° and cell volume V= 223.8626 A°$^3$ ). These results are similar to those observed in the nanowires of $Pr_{0.5}Ca_{0.5}MnO_3$ [13] . The SEM images (Fig.2) of $Pr_{0.57}Ca_{0.41}Ba_{0.02}MnO_3$ nanowires show that the sample consists of a large number of nanowires randomly distributed one over the other there being no preferential orientation of the nanowires. The TEM images (Fig. 3) also confirm the wire nature of the sample with lengths in a few microns range. The bright field image of a single nanowire is shown in Fig. 3(a). The diameter of the wire is in the range of about 80-90 nm. In addition to the nanowires some particles were also observed which had different grain sizes in the range of 20-40 nm. Interestingly some particles were hexagonal in shape and are also much larger as shown in Fig. 3(b). Fig. 4 shows the field cooled magnetization of undoped $Pr_{0.57}Ca_{0.43}MnO_3$ (bulk) along with those of Ba doped bulk and nanowires of $Pr_{0.57}Ca_{0.41}Ba_{0.02}MnO_3$, recorded in the temperature range 10-300 K in the presence of a magnetic field 0.1 T. It is seen that in the undoped material only the CO transition is observed with the magnetization peak around 245 K. In the Ba doped material a ferromagnetic phase develops at low temperatures (the transition temperature $T_C$ = 110 K as determined by a minimum in dM/dT). In the naowires, however, the charge order that was seen in the bulk sample is completely absent with only the ferromagnetic transition occurring at $T_C$ = 115 K is observed. The $T_C$ for the nanowires was slightly higher than that of the bulk though the magnetization was less compared to that of the bulk. EPR signals (not shown) at room temperature for both the bulk and nano PCBM are broad single lines with the peak to peak widths of 1040 Oe and 1437 Oe respectively. The observed larger EPR line width of the nanowire sample, which is attributable to the disordered surface layer appears to be a generic feature and could serve as a tool for quick identification of the nanometric systems.

In the following we attempt to provide a qualitative explanation for the most significant result of the present work, namely, the complete disappearance of the charge order in the nanowires. Indeed, it is no surprise that the properties of nanosystems could be very different from those of the bulk since the combination of the increased surface to volume ratio and finite size effects could have profound effect on the behaviour of nanosystems. Earlier workers have invoked various other mechanisms such as a reduction in the unit cell size and anisotropy to explain the enhancement of $T_C$ observed in the nanosystems. That explanation is clearly not applicable in the present case since we do not observe any reduction either in the size or in the anisotropy of the unit cell. In very general terms, ordering at lower dimensions is suppressed due to the stronger fluctuations in smaller systems. For the present, we offer this qualitative explanation for the absence of charge order in the nanowires and nanoparticles, though clearly more work is needed to substantiate this proposal. In contrast, the fact that the nanosystems are ferromagnetic at low temperatures could be likely due to the much shorter range of the mediating interaction (i.e. exchange) in the case of magnetism. Incidentally, the difference in the magnetization observed in the nanowires compared to the bulk could quite simply be due to the presence of the larger contribution of the surface shell with spin disorder.

KNA acknowledges a research fellowship from the All India Council for Technical education.

* **Corresponding author**

**Figure Captions:**

Fig. 1 Observed (dots) and Rietveld fitted (continuous lines) powder X-ray diffraction pattern of $Pr_{0.57}Ca_{0.41}Ba_{0.02}MnO_3$ nanowires.

Fig. 2 Typical SEM images of $Pr_{0.57}Ca_{0.41}Ba_{0.02}MnO_3$ nanowires.

Fig. 3 Typical TEM images of $Pr_{0.57}Ca_{0.41}Ba_{0.02}MnO_3$ nanowires. (a) bright field image of a single nanowire of diameter approximately 80nm (b) a hexagonal particle is shown in between two nanowires.

Fig. 4 Temperature dependence of magnetization for (a) bulk $Pr_{0.57}Ca_{0.43}MnO_3$ showing $T_{CO} = 245$ K (b) bulk $Pr_{0.57}Ca_{0.41}Ba_{0.02}MnO_3$ showing $T_{CO} = 230$ K and $T_C = 110$ K (c) Nanowires of $Pr_{0.57}Ca_{0.41}Ba_{0.02}MnO_3$ showing no charge order but $T_C = 115$ K. Inset shows a clear picture of disappearance of charge order in nano wires.

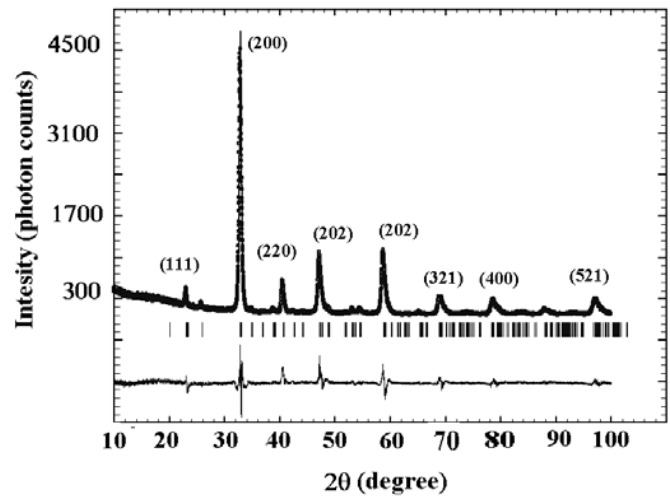

Figure 1 : K.N.Anuradha , S.S.Rao and S.V.Bhat

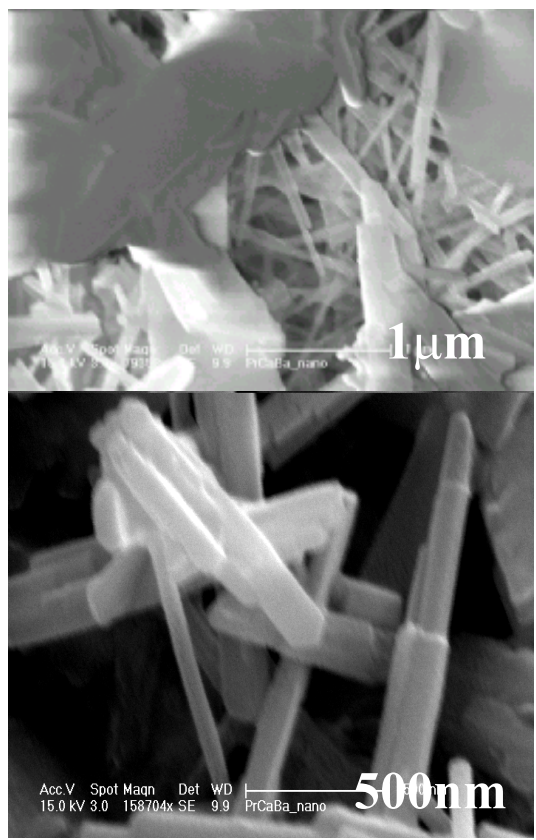

Figure 2: K.N.Anuradha , S.S.Rao and S.V.Bhat

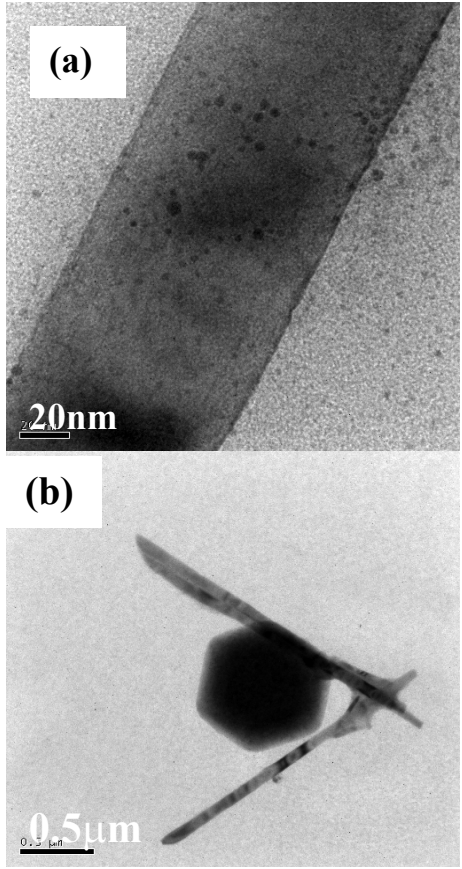

Figure 3: K.N.Anuradha , S.S.Rao and S.V.Bhat

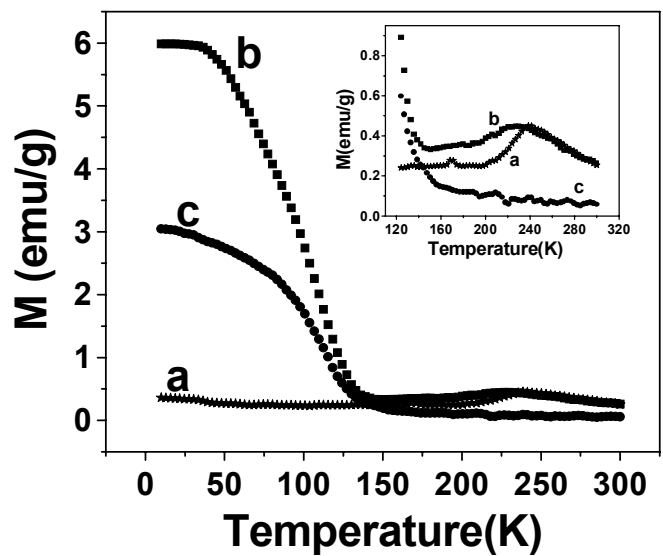

Figure 4: K.N.Anuradha , S.S.Rao and S.V.Bhat